\documentclass[prd,a4paper,showpacs,byrevtex]{revtex4}
\usepackage{graphicx}
\usepackage{dcolumn}
\usepackage{amsmath}
\usepackage{array}
\usepackage{bm}
\usepackage{amssymb}
\usepackage{amsfonts}

\begin{document}

\title{Gluons in glueballs: Spin or helicity?}

\author{Vincent \surname{Mathieu}}
\thanks{IISN Scientific Research Worker}
\email[E-mail: ]{vincent.mathieu@umh.ac.be}
\author{Fabien \surname{Buisseret}}
\thanks{F.R.S.-FNRS Research Fellow}
\email[E-mail: ]{fabien.buisseret@umh.ac.be}
\author{Claude \surname{Semay}}
\thanks{F.R.S.-FNRS Research Associate}
\email[E-mail: ]{claude.semay@umh.ac.be}
\affiliation{Groupe de Physique Nucl\'{e}aire Th\'{e}orique,
Universit\'{e} de Mons-Hainaut,
Acad\'{e}mie universitaire Wallonie-Bruxelles,
Place du Parc 20, BE-7000 Mons, Belgium}

\date{\today}

\begin{abstract}
In the last decade, lattice QCD has been able to compute the low-lying glueball spectrum with accuracy. Like other effective approaches of QCD, potential models still have difficulties to cope with gluonic hadrons. Assuming that glueballs are bound states of valence gluons with zero current mass, it is readily understood that the use of a potential model, intrinsically non covariant, could be problematic in this case. The main challenge for this kind of model is actually to find a way to introduce properly the more relevant degree of freedom of the gluon: spin or helicity. In this work, we use the so-called helicity formalism of Jacob and Wick to describe two-gluon glueballs. We show in particular that this helicity formalism exactly reproduces the $J^{PC}$ numbers which are observed in lattice QCD when the constituent gluons have a helicity-1, without introducing extra states as it is the case in most of the potential models. These extra states appear when gluons are seen as spin-1 particles. Using a simple spinless Salpeter model with Cornell potential within the helicity formalism, we obtain a glueball mass spectrum which is in good agreement with lattice QCD predictions for helicity-1 gluons provided instanton-induced interactions are taken into account.
\end{abstract}



\maketitle

\section{Introduction}

As quantum chromodynamics (QCD) is built on the nonabelian SU(3)-color group, it allows for purely gluonic bound states called glueballs. The structure and properties of
these pure glue states is nowadays far from being completely understood and deserves much interest on
both experimental and theoretical sides. 

On the one hand, some experimental glueball candidates are currently known. Most of them are scalar,
such as the $a_0(980)$, $f_0(980)$, $f_0(1500)$, $f_0(1710)$, \dots but no definitive
conclusions can be drawn concerning the nature of these states. Indeed, it is often pointed out that the
lightest glueballs are probably strongly mixed with other hadrons like mesons and tetraquarks for example. Many details concerning the identification of experimental states can be found in the recent report~\cite{exp_gg}. 

On the other hand, pure gauge QCD has been investigated by lattice QCD for many years, leading to a well established glueball spectrum below 4~GeV \cite{lat0,lat1}. Numerous effective QCD models have also been applied to study the glueballs. One can quote QCD sum rules~\cite{sumrules,inst2}, AdS/QCD correspondence~\cite{AdS}, Coulomb gauge QCD~\cite{cg2,szcz}, and potential models. Pioneering works in this last field were presented in Refs.~\cite{bar,corn}. In both works, glueballs are seen as bound states of at least two valence gluons, but the properties of these gluons (mass and spin) are very different. In the first reference, valence gluons are assumed to be helicity-1 particles. It means that their spin has only two projections ($\pm 1$). In the second one however, they are seen as massive particles with spin-1 (the zero projection is also allowed).  

A remark should here be done concerning the number of constituent gluons in a given glueball. It appears in lattice QCD that the lowest-lying glueballs are the $C=+$ ones. As a bound state of two gluons can only have $C=+$, it is rather natural to assume that the lightest glueballs are mainly two-gluon states (the more constituent gluons are present, the more the glueball should be heavy). This picture, that we adopt in the present work, is widely accepted in models with constituent gluons. Moreover, it is interesting to mention some results of the Coulomb gauge study of Ref.~\cite{szcz}. In this approach, a Fock space expansion of glueball states in terms of quasigluons can be performed, and it appears that the influence of the three- and four-gluon components on the low-lying $C=+$ glueballs is negligible: The two-gluon component is dominant as intuitively expected. But actually, the relevance of using a potential model to describe a glueball is still controversial. Let us begin by the problem of the gluon mass. As we already mentioned, there are works, in the spirit of Ref.~\cite{bar}, arguing that a valence gluon is a massless particle, which gains a constituent mass $\mu$, either constant~\cite{bar}, or state-dependent $\mu=\left\langle \sqrt{\vec p^{\, 2}}\right\rangle$~\cite{simo2,brau,buisnew}. Relativistic spin-dependent corrections are then developed in powers of $1/\mu^2$. In this picture, the valence gluon is \textit{a posteriori} massive, because it is confined into a glueball. Let us note that, more generally, both quarks and gluons can gain a constituent mass from renormalization theory (in the Coulomb gauge approach of Ref.~\cite{llan1}, massless gluons gain a constituent mass of about $0.7$~GeV at zero momentum). But, other studies keep the assumptions of Ref.~\cite{corn} and state that a valence gluon has to be \textit{a priori} considered as massive, that is with a non zero current mass \cite{corn,hou}. The underlying idea is roughly that the nonperturbative effects of QCD cause a mass term to appear in the gluon propagator. This fixed gluon mass is typically assumed to be around $m_g=0.5\pm0.2$~GeV \cite{Bernard:1981pg,Bonnet:2000kw,aguil}, and the relativistic corrections are then expanded in powers of $1/m^2_g$. The problem of the gluon spin is obviously linked to its mass. If a valence gluon is a priori massive, then it is a spin-1 particle. But, if it is massless, what is the correct internal degree of freedom? The most obvious answer is that it has a helicity-1. But, the spin corrections in potential models appear at the order $1/\mu^2$, at a level where the gluons have a dynamical mass, and thus perhaps a spin-1. In the present work, we will only focus on the case where the valence gluons are massless, that is have a vanishing current mass. We thus need a formalism which allows us to deal with both helicity and spin degrees of freedom, and to build quantum states with the correct symmetry following the degree of freedom which is chosen. 

In potential approaches, hadrons are generally described by $\left|^{2S+1}L_J\right\rangle$
quantum states (in spectroscopic notation) which are simultaneously eigenstates of $\vec J^{\, 2}$, $\vec L^{\, 2}$ and $\vec S^{\,2}$, such that $\vec J=\vec L+\vec S$ is the total spin. But actually, $\vec J$ is the only relevant angular momentum labeling a hadronic state (together with the parity and the charge conjugation). In this
picture, it is assumed that a $J^{PC}$ state is a linear combination of the allowed $\left|^{2S+1}L_J\right\rangle$ states leading to the desired value of the total spin. How to build such a general state? We propose here to use the helicity formalism, developed by Jacob and Wick in Ref.~\cite{jaco} to describe scattering in two-body systems. The crucial feature of this very general formalism is that it remains valid for massless particles like gluons. But, as we will show through this paper, it provides a powerful tool to build $J^{PC}$ states in terms of the usual $\left|^{2S+1}L_J\right\rangle$ states, and allows to gain considerable insight on the glueball models in potential approaches. 

Our paper is organized as follows. Sec.~\ref{genform} is a presentation of the helicity formalism. It sums up the key points of Ref.~\cite{jaco}. Then, in Sec.~\ref{hsglu}, this formalism, which allows for the gluons to have either a helicity or a spin degree of freedom, is applied to two-gluon
glueballs. The case of massless gluons has already been studied in Ref.~\cite{bar}. We
present here a more detailed study with a proper treatment of the relativistic kinematics, and we formulate the glueball
helicity states in a way that is more convenient to further apply
to potential models. Moreover, the glueball spectrum is now far
better known than at the time of Ref.~\cite{bar} thanks to lattice
QCD calculations~\cite{lat0,lat1}. That is why it is of interest to
reconsider the description of glueballs with helicity states. To this aim, we introduce in Sec.~\ref{pmodel} a simple potential model based on a spinless Salpeter Hamiltonian with a Cornell potential. Instanton-induced forces are also included. Then, we show in Sec.~\ref{numapp} that the model we introduced, supplemented by the helicity state formalism for helicity-1 gluons, leads to a rather good agreement with the lattice QCD spectrum. Finally,
we draw some conclusions in Sec.~\ref{conclu}.

\section{Helicity formalism}\label{genform}
\subsection{General considerations}

Let $\left|\psi_{k,\lambda}\right\rangle$ be the state of a particle of momentum $\vec k=k\vec 1_z$ and helicity $\lambda$. For a particle of spin $s$ and mass $m$, one can have
\begin{eqnarray}
\lambda&=&-s,-s+1,\dots,+s\quad   (m\neq 0),\\
&&\pm s \hspace{2.78cm}  (m=0).
\end{eqnarray}
If we define a general rotation as $R(\alpha,\beta,\gamma)=\exp(-i\alpha J_z)\, \exp(-i\beta J_y)\, \exp(-i\gamma J_z)$, with $\alpha,\beta,\gamma$ the Euler angles and $\vec J$ the angular momentum operators for the considered particle, then
\begin{equation}\label{1pstate}
    \left|\psi_{p,\lambda}\right\rangle=R(\phi,\theta,-\phi)\, \left|\psi_{k,\lambda}\right\rangle={\rm e}^{i\lambda\phi}R(\phi,\theta,0)\, \left|\psi_{k,\lambda}\right\rangle
\end{equation}
is the state of a particle whose helicity is $\lambda$ and whose momentum $\vec p$ $(\left|\vec p\, \right|=k)$ has the arbitrary polar angles $(\theta,\phi)$.

In the reference frame where $\vec p_1=-\vec p_2=\vec p$, a two-particle state $\left|\psi_{p,\lambda_1,\lambda_2}\right\rangle$ can be built from $\left|\psi_{p,\lambda_1}\right\rangle$ and $\left|\psi_{p,\lambda_2}\right\rangle$, both given by Eq.~(\ref{1pstate}). It reads
\begin{equation}\label{2pstate}
    \left|\psi_{p,\lambda_1,\lambda_2}\right\rangle=\left|\psi_{p,\lambda_1}\right\rangle\otimes\left|\chi_{p,\lambda_2}\right\rangle,
\end{equation}
with
\begin{equation}
    \left|\chi_{p,\lambda_2}\right\rangle=(-1)^{s_2-\lambda_2}{\rm e}^{-i\pi J^{(2)}_y}\left|\psi_{p,\lambda_2}\right\rangle.
\end{equation}
The rotation along the $y$ axis ensures that $\vec p_1=-\vec p_2$, while the phase factor $(-1)^{s_2-\lambda_2}$ is such that $ \left|\chi_{0,\lambda}\right\rangle=\left|\psi_{0,-\lambda}\right\rangle$ as intuitively expected.

As it is defined in Eq.~(\ref{2pstate}), the two-particle state is not an eigenstate of the square total angular momentum $\vec J^{\, 2}$ and of its projection $M(=J_z)$. However, the state
\begin{equation}\label{hstate}  \left|J,M;\lambda_1,\lambda_2\right\rangle=\left[\frac{2J+1}{4\pi}\right]^{1/2}\int^{2\pi}_0d\phi\int^\pi_0d\theta\, \sin\theta\ {\cal D}^{J*}_{M,\lambda_1-\lambda_2}(\phi,\theta,-\phi)\, R(\phi,\theta,-\phi)\,  \left|\psi_{p,\lambda_1,\lambda_2}\right\rangle
\end{equation}
is, by construction, an eigenstate of $\vec J^{\, 2}$ and $J_z$ (more details can be found in Ref.~\cite[Chapter 7]{taylor} for example). Indeed, the Wigner $D$-functions
\begin{equation}
    {\cal D}^{J}_{M,M'}(\alpha,\beta,\gamma)={\rm e}^{-iM\alpha}\, d^J_{M,M'}(\beta)\, {\rm e}^{-iM'\gamma}
\end{equation}
enforce a particular value for the total angular momentum. Their explicit forms can be found for example in Ref.~\cite[Chapter 4]{var}. The states given by Eq.~(\ref{hstate}) are orthonormalized by definition, i.e. $\left\langle J',M';\lambda'_1,\lambda'_2\right|\left.J,M;\lambda_1,\lambda_2\right\rangle=\delta_{J,J'}\delta_{M,M'}\delta_{\lambda_1,\lambda'_1}\delta_{\lambda_2,\lambda'_2}$, and they describe a general two-particle system in the rest frame. The use of helicity degrees of freedom rather than the spin ones allows to deal with massless particles too. This feature will obviously be useful in the description of glueballs. Let us note that we did not write explicitly the dependence in $p$ of the helicity states in order to simplify the notation.


\subsection{Symmetries}
The states~(\ref{hstate}) are defined so that they satisfy
\begin{eqnarray}
\vec J^{\, 2}\left|J,M;\lambda_1,\lambda_2\right\rangle&=&J(J+1)\left|J,M;\lambda_1,\lambda_2\right\rangle, \\
    J_z\left|J,M;\lambda_1,\lambda_2\right\rangle&=&M\left|J,M;\lambda_1,\lambda_2\right\rangle.
\end{eqnarray}
Moreover, the usual rules concerning the coupling of two angular momenta lead to the constraint
\begin{equation}\label{jinf}
    J\geq\left|\lambda_1-\lambda_2\right|.
\end{equation}
It can also be shown that the $\left|J,M;\lambda_1,\lambda_2\right\rangle$ states have the following behavior under the parity $\hat P$:
\begin{equation}\label{pdef}
    \hat P\left|J,M;\lambda_1,\lambda_2\right\rangle=\eta_1\eta_2(-1)^{J-s_1-s_2}\left|J,M;-\lambda_1,-\lambda_2\right\rangle,
\end{equation}
where $\eta_i$ is the intrinsic parity of particle $i$. A physical state is asked not only to be an eigenstate of the total angular momentum operators but also of the parity. Such a requirement is fulfilled by the following linear combinations
\begin{eqnarray}\label{hdef0}
    \left|H_\pm; J^P;\lambda_1,\lambda_2\right\rangle&=&\frac{1}{\sqrt 2}\left\{\, \left|J,M;\lambda_1,\lambda_2\right\rangle\pm\left|J,M;-\lambda_1,-\lambda_2\right\rangle\right\}, \quad \lambda_1\ {\rm or}\ \lambda_2\neq0,\\
    \left|N; J^P\right\rangle&=& \left|J,M;0,0\right\rangle,
\end{eqnarray}
for which $\hat P   \left|H_\pm; J^P;\lambda_1,\lambda_2\right\rangle=P \left|H_\pm; J^P;\lambda_1,\lambda_2\right\rangle$, with $P=\pm\eta_1\eta_2(-1)^{J-s_1-s_2}$. In the latter, the $\left|H_\pm; J^P;\lambda_1,\lambda_2\right\rangle$ and $\left|N; J^P\right\rangle$ states will be referred as helicity states. 

When the two particles are identical ($m_1=m_2=m$, $s_1=s_2=s$), it is relevant to study the action of the permutation operator $P_{12}$. One finds
\begin{equation}\label{symdef}  \left[1+(-1)^{2s}P_{12}\right]\left|J,M;\lambda_1,\lambda_2\right\rangle=\left|J,M;\lambda_1,\lambda_2\right\rangle+(-1)^{J}\left|J,M;\lambda_2,\lambda_1\right\rangle,
\end{equation}
  where the operator $\left[1+(-1)^{2s}P_{12}\right]=\hat {\cal S}$ is nothing else than a projector on the symmetric ($s$ integer) or antisymmetric ($s$ half-integer) part of the helicity state. Consequently, in the special case of identical particles, the helicity states should also be eigenstates of $\hat {\cal S}$. By inspection of relations~(\ref{hdef}) and (\ref{symdef}), it can be seen that the states 
\begin{eqnarray}\label{hdef}
	\left|H_{\pm,\rho};J^P\right\rangle&=&\frac{1}{\sqrt{2}}\left\{\left|H_\pm; J^P;\lambda_1,\lambda_2\right\rangle+\rho\left|H_\pm; J^P;\lambda_2,\lambda_1\right\rangle\right\}, \quad \lambda_1\ {\rm or}\ \lambda_2\neq0,\\
	 \left|N; J^P\right\rangle&=& \left|J,M;0,0\right\rangle,
\end{eqnarray}
 with $\rho=\pm1$, are eigenstates of $\hat{\cal S}$. More precisely,
  \begin{subequations}\label{selecgen}
\begin{eqnarray}
    \hat{\cal S}\left|H_{\pm,\rho}; J^P\right\rangle&=&\left[1+\rho(-1)^J\right]\left|H_{\pm,\rho}; J^P\right\rangle ,\\
    \hat{\cal S}\left|N; J^P\right\rangle&=&\left[1+(-1)^J\right]\left|N; J^P\right\rangle .
\end{eqnarray}
\end{subequations}

In the symmetric case, we thus observe the emergence of selection rules following the value of $J$ and $\rho$. 

\subsection{Wave functions}

It is of great phenomenological interest to be able to express a given helicity state in terms of states of given orbital angular momentum $L$ and intrinsic spin $S$. Indeed, although the total spin $J$ is the only relevant angular momentum of the system, especially when one deals with relativistic bound states, most of the Hamiltonian-based effective approaches of QCD involve central potentials with relativistic spin corrections. Such Hamiltonians thus act on nonrelativistic $\left|^{2S+1}L_J\right\rangle$ states (in spectroscopic notation) rather than on helicity states. It is actually proved in Ref.~\cite{jaco} that the following decomposition holds:
\begin{equation}\label{decomp}
    \left|J,M;\lambda_1,\lambda_2\right\rangle=\sum_{L,S}\left[\frac{2L+1}{2J+1}\right]^{1/2} \left\langle L,S;0,\lambda_1-\lambda_2\right|J,\lambda_1-\lambda_2\left.\right\rangle\left\langle s_1,s_2;\lambda_1,-\lambda_2\right|S,\lambda_1-\lambda_2\left.\right\rangle\, \left|^{2S+1}L_J\right\rangle,
\end{equation}
where we also impose the normalization condition
\begin{equation}
    \left\langle^{2S'+1}L'_{J'}\left|\right.^{2S+1}L_J\right\rangle=\delta_{L,L'}\, \delta_{S,S'}\, \delta_{J,J'}.
\end{equation}
The sum~(\ref{decomp}) involves all the $\{ L,S\}$ couples such that $\vec S=\vec s_1+\vec s_2$ and $\vec L+\vec S=\vec J$. The symbols $\left\langle a,b;c,d\right|e,f\left.\right\rangle$ denote the well-known Clebsch-Gordan coefficients.

More explicitly, one can write the $\left|^{2S+1}L_J\right\rangle$ states as
\begin{equation}\label{lsdef}
    \left|^{2S+1}L_J\right\rangle=\left|R_J(r)\right\rangle\otimes\left[\left|Y^{L}(\hat r)\right\rangle\otimes\left|s_1,s_2\right\rangle^S\right]^J,
\end{equation}
where the radial, angular, and spin wave functions are explicitly written. It is worth noting that the shape of the radial wave function can only depend on the total angular momentum $J$ in order not to destroy the symmetry properties of the helicity states. At this stage, one could wonder if the helicity formalism, which has been proposed as a powerful way of studying relativistic scattering problems, can be applied to describe bound states. Actually, the construction of the helicity states that was presented here is purely geometrical. The angular parts of the states are built in order to have the desired properties, but the radial part of each state is arbitrary (excepted that it can only depend on $J$). Consequently, only the dynamics of the system will fix this radial part: Spherical waves will be obtained for scattering states, and bound state radial wave functions otherwise. But, the construction we presented up to now, being Hamiltonian-independent, is valid in both cases.

\section{Helicity states for two-gluon glueballs}\label{hsglu}

\subsection{Gluons with helicity}

We now particularize the formalism to the special case of a system made of two gluons with helicity-1. Then, $m_1=m_2=0$, $s_1=s_2=1$, $\lambda_i=\pm1$, $\eta_1\eta_2=1$. Actually, the results that we will obtain in this section are formally identical to the case of a state made of two photons. The total color wave function is indeed assumed to be a singlet one, which is totally symmetric, and does not explicitly appear in the computations. Taking into account the symmetrization of the helicity sates, one finds that there are four allowed helicity states, namely
\begin{equation}\label{hsdef}
    \left|S_\pm;J^P\right\rangle=\left| H_{\pm,1}; J^P\right\rangle_{\lambda_2=\lambda_1}, \quad \left|D_+;J^P\right\rangle=\left| H_{+,1}; J^P\right\rangle_{\lambda_2=-\lambda_1}, \quad \left|D_-;J^P\right\rangle=\left| H_{-,-1}; J^P\right\rangle_{\lambda_2=-\lambda_1}.
\end{equation}

But, the selection rules~(\ref{selecgen}) together with Eqs.~(\ref{jinf}) and (\ref{pdef}) impose particular values for the total angular momentum and parity of these four states. It can be checked that one can only obtain the following states
\begin{equation}\label{ggstate}
        \left|S_+;(2k)^+\right\rangle,\quad \left|S_-;(2k)^-\right\rangle,\quad \left|D_+; (2k+2)^+\right\rangle,\quad \left|D_-; (2k+3)^+\right\rangle,\quad \quad k\in\mathbb{N}.
\end{equation}
The $S$- and $D$-labels stand for helicity-singlet and -doublet respectively.

It should be noticed that a state made of two gluons in a color singlet has always a positive charge conjugation ($C=+1$). More explicitly, the states~(\ref{ggstate}) thus give rise to the following glueball states
\begin{subequations}\label{ggstate3}
\begin{eqnarray}
    \left|S_+;(2k)^+\right\rangle &\Rightarrow& 0^{++}, 2^{++}, 4^{++},\dots\\
    \left|S_-;(2k)^-\right\rangle&\Rightarrow&0^{-+}, 2^{-+}, 4^{-+},\dots\\
    \left|D_+;(2k+2)^+\right\rangle&\Rightarrow& 2^{++},4^{++},\dots\\
        \left|D_-;(2k+3)^+\right\rangle&\Rightarrow&3^{++},5^{++},\dots
\end{eqnarray}
\end{subequations}
It is readily observed that only the $\left|S_\pm;(2k)^+\right\rangle$ states can lead to $J=0$, while the $\left|D_\pm\right\rangle$ states always have $J\geq 2$. Obviously, no $J=1$ state is present: Only the $\left|D_-\right\rangle$ states can generate an odd-$J$, but $J$ is at least 3 in this case. The fact that a state made of two photons (or gluons) can never have the value $J=1$ is known as Yang's theorem~\cite{yang}, but has also been found independently by Landau~\cite{landau}. Lattice QCD confirm the absence of $1^{-+}$ and $1^{++}$ states, at least below $4$ GeV. It is worth mentioning that glueball states with an even-$J$ and a positive parity can be built either from the helicity-singlet or from the helicity-doublet. The important fact is that the helicity states exactly reproduce the $J^{PC}$ glueballs which are observed in lattice QCD without the extra states which are usually present in potential models.

The application of the decomposition formula~(\ref{decomp}) to the helicity states~(\ref{ggstate}) gives
\begin{subequations}\label{ggstate2}
\begin{eqnarray}
    \left|S_+;(2k)^+\right\rangle&=&\left[\frac{2}{3}\right]^{1/2}\left|^1 2k_{2k}\right\rangle-\left[\frac{2k(2k+1)}{3(4k-1)(4k+3)}\right]^{1/2}\left|^52k_{2k}\right\rangle\nonumber\\
    &&+\left[\frac{k(2k-1)}{(4k+1)(4k-1)}\right]^{1/2}\left|^5 2k-2_{2k}\right\rangle+\left[\frac{(k+1)(2k+1)}{(4k+3)(4k+1)}\right]^{1/2}\left|^5 2k+2_{2k}\right\rangle,\\
    \left|S_-;(2k)^-\right\rangle&=&\left[\frac{2k}{4k+1}\right]^{1/2}\left|^3 2k-1_{2k}\right\rangle-\left[\frac{2k+1}{4k+1}\right]^{1/2}\left|^32k+1_{2k}\right\rangle,\\
    \left|D_+;(2k+2)^+\right\rangle&=&\left[\frac{(k+2)(2k+3)}{(4k+3)(4k+5)}\right]^{1/2}\left|^52k_{2k+2}\right\rangle
    +\left[\frac{6(k+2)(2k+1)}{(4k+3)(4k+7)}\right]^{1/2}\left|^5 2k+2_{2k+2}\right\rangle\nonumber\\
    &&+\left[\frac{(k+1)(2k+1)}{(4k+5)(4k+7)}\right]^{1/2}\left|^5 2k+4_{2k+2}\right\rangle,\\
        \left|D_-;(2k+3)^+\right\rangle&=&-\left[\frac{2k+5}{4k+7}\right]^{1/2}\left|^52k+2_{2k+3}\right\rangle
    -\left[\frac{2(k+1)}{4k+7}\right]^{1/2}\left|^5 2k+4_{2k+3}\right\rangle.\label{dms}
\end{eqnarray}
\end{subequations}

\begin{table}[ht]
\caption{Matrix elements of a given operator $\hat{\cal O}$ for the glueball states composed of two helicity-1 gluons. All non diagonal elements are vanishing. The minimal $J^{PC}$ values are indicated in parenthesis.}
\label{tab1}
\begin{ruledtabular}
\begin{tabular}{ccccccc}
$\hat{\cal O}$ & $\textbf{1}$ &$\vec L^{\, 2}$ & $\vec S^{\, 2}$ & $\vec L\cdot\vec S$ & $\vec J^{\, 2}$ & $\vec S^{\, 2}-3(\vec S\cdot \hat r)^2$ \\
\hline
$\left|S_+ ;J^+\right\rangle$ ($0^{++}$)   &1& $J(J+1)+2$ & 2 & $-2$ & $J(J+1)$ & 2\\
$\left|S_- ;J^-\right\rangle$ ($0^{-+}$)   &1& $J(J+1)+2$ & 2 & $-2$ & $J(J+1)$ &2\\
$\left|D_+ ;J^+\right\rangle$ ($2^{++}$)   &1& $J(J+1)-2$ & 6 & $-2$ & $J(J+1)$ &$-6$\\
$\left|D_- ;J^+\right\rangle$ ($3^{++}$)   &1& $J(J+1)-2$ & 6 & $-2$ & $J(J+1)$ &$-6$\\
\end{tabular}
\end{ruledtabular}
\end{table}

Thanks to the decompositions~(\ref{ggstate2}), the matrix elements of various operators are readily computed. They are given in Table~\ref{tab1}. Note that, through the kinetic energy, the $\vec L^{\, 2}$ operator controls the glueball mass in a simple Hamiltonian with only a central potential. It appears that the matrix elements only depend on the singlet or doublet nature of an helicity state. Moreover, the matrix elements between the various even-$J^{++}$ states, $\left|S_+ ;(J\geq2)^+\right\rangle$  and $\left|D_+ ;(J\geq2)^+\right\rangle$, vanish. The helicity-singlet and helicity-doublet are thus completely decoupled. It was claimed in Ref.~\cite{bar} that the glueball spectrum should be characterized by a tower of degenerate even-$J$ glueball states with positive and negative parity (the helicity-singlets). However, recent lattice QCD computations have unambiguously shown that the $0^{++}$ and $0^{-+}$ glueballs are not degenerate, as well as the $2^{++}$ and $2^{-+}$ ones~\cite{lat0,lat1}. The difference between the $2^{++}$ and $2^{-+}$ states can be easily explained because the lightest $2^{++}$ states should be an helicity doublet, and not an helicity singlet as the $2^{-+}$. In addition there should be a
$2^{++}$ helicity singlet with the same mass than the $2^{-+}$ glueball, as pointed out in Ref.~\cite{bar}. This has not been detected in lattice QCD. The problem of the degeneracy of the $0^{-+}$ and $0^{++}$ states however, requires a particular Hamiltonian to be elucidated. One could think for example to instanton-induced interactions, which are repulsive in the pseudo-scalar channel, and attractive in the scalar channel~\cite{inst,inst2,Kochelev:2005vd}. Such interactions, that we will further comment in Sec.~\ref{ins}, will lead to a correct ordering of the glueball states.

\subsection{Gluons with spin}\label{gspin}

It is also possible that the gluons should be considered as spin-1 particles rather than helicity-1 ones. The most obvious way to have this situation is to deal with massive gluons. It should be stressed that, in every glueball potential model, the relativistic corrections (containing the spin-dependent terms) involve a constituent gluon mass. This is true even if the gluon is massless at the dominant order. Consequently, it could be possible, in the framework of effective models, to deal with massless gluons at the dominant order but to give them a spin degree of freedom because they are massive at the order of the spin-dependent terms. If the gluons have spin-$1$ rather than helicity-1 the value $\lambda_i=0$ can be reached. However, one is always dealing with identical bosons such that $m_1=m_2$ and $\eta_1=\eta_2$. Consequently, as it can be deduced from relations~(\ref{selecgen}), the helicity states describing a glueball made of two gluons with spin are the four states~(\ref{ggstate2}) supplemented by
\begin{subequations}
\begin{eqnarray}\label{nstate}
    \left|N; (2k)^+\right\rangle&\Rightarrow& 0^{++},\, 2^{++},\, 4^{++},\dots,\\
    \left|H_{+,1};(2k+2)^+\right\rangle_{\lambda_1=0,\lambda_2=1}&\Rightarrow& 2^{++},\, 4^{++},\, 6^{++},\dots\\  
    \left|H_{-,1};(2k+2)^+\right\rangle_{\lambda_1=0,\lambda_2=1}&\Rightarrow& 2^{-+},\, 4^{-+},\, 6^{-+},\dots\\ 
     \left|H_{+,-1};(2k+1)^-\right\rangle_{\lambda_1=0,\lambda_2=1}&\Rightarrow& 1^{-+},\, 3^{-+},\, 5^{-+},\dots\\ 
      \left|H_{-,-1};(2k+1)^+\right\rangle_{\lambda_1=0,\lambda_2=1}&\Rightarrow& 1^{++},\, 3^{++},\, 5^{++},\dots
\end{eqnarray}
\end{subequations}
Five additional states appear because of the allowed zero value for the helicity. First, we can point out the apparition of a family of glueballs with odd $J$ and negative parity. Such states are actually not observed in lattice QCD. The decomposition formula~(\ref{decomp}) leads to $ \left|H_{+,-1};(2k+1)^-\right\rangle_{\lambda_1=0,\lambda_2=1}=\left|^3J_J\right\rangle$, i.e. a pure $\left|^{2S+1}L_J\right\rangle$ state. 

It is also worth mentioning that a $1^{++}$ glueball, forbidden with helicity-1 gluons, is now allowed. Applying Eq.~(\ref{decomp}), one finds that 
\begin{equation}\label{nstate2} \left|H_{-,-1};(2k+1)^+\right\rangle_{\lambda_1=0,\lambda_2=1}=\left[\frac{2k}{4k+3}\right]^{1/2}\left|^5 2k_{2k+1}\right\rangle-\left[\frac{2k+3}{4k+3}\right]^{1/2}\left|^5 2k+2_{2k+1}\right\rangle, 
\end{equation}
that is a decomposition which is very similar to the helicity-doublet~(\ref{dms}), which generates the same quantum numbers for $J\geq3$. Both states can actually couple to each other through the orbital angular momentum: $\left\langle H_{-,-1};J^+\right|\vec L^{\, 2}\left|D_-;J^+\right\rangle=2\sqrt{(J-1)(J+2)}$. Let us further investigate this point by considering the dynamics of the system through its Hamiltonian $\hat H$. As the $\left|D_-;J^+\right\rangle$ and $\left|H_{-,-1};J^+\right\rangle$ states are coupled, the physical states are eigenvectors of the Hamiltonian 
\begin{equation}\label{ham_mat_g}
{\cal H}=\begin{pmatrix}
\left\langle D_-;J^+\right|\hat H\left|D_-;J^+\right\rangle & \left\langle D_-;J^+\right|\hat H\left|H_{-,-1};J^+\right\rangle\\
\left\langle H_{-,-1};J^+\right|\hat H\left|D_-;J^+\right\rangle & \left\langle H_{-,-1};J^+\right|\hat H\left|H_{-,-1};J^+\right\rangle
\end{pmatrix}   ,
\end{equation}
where these four matrix elements can be expressed as linear combinations of $\left\langle ^{2S'+1}L'_J\right|\hat H\left|^{2S+1}L_J\right\rangle$ thanks to relations~(\ref{ggstate2}) and (\ref{nstate2}). Then, it can be shown that the eigenstates of Hamiltonian~(\ref{ham_mat_g}) are the pure $\left|^5 J-1_J\right\rangle$ and $\left|^5 J+1_J\right\rangle$ states. This result is obtained under the assumption that no tensor force is present. This is always the case at the dominant order, in particular when one deals with central potentials with relativistic corrections.   

In the same way, assuming that $\left\langle ^{2S'+1}L'_J\right|\hat H\left|^{2S+1}L_J\right\rangle \propto \delta_{S',S}\delta_{L',L}$, one can check that the coupling between the $\left|S_-;J^+\right\rangle$ and $\left|H_{-,1};J^+\right\rangle_{\lambda_1=0,\lambda_2=1}$ states for $J\geq 2$ leads to the conclusion that $\left|^3J-1_J\right\rangle$ and $\left|^3 J+1_J\right\rangle$ are the quantum states that have to be considered. Finally, the $\left|N; J^+\right\rangle$ and $\left|H_{+,1};J^+\right\rangle_{\lambda_1=0,\lambda_2=1}$ and helicity states are coupled to each other and to the $\left|S_+;J^+\right\rangle$ and $\left|D_+;J^+\right\rangle$ states, so that we have checked that the physical quantum states are  $\left|^1J_J\right\rangle$, $\left|^5J-2_J\right\rangle$, $\left|^5J_J\right\rangle$, and $\left|^5J+2_J\right\rangle$.   

In conclusion, when the dynamics of the system is included, the nine possible helicity states reduce to the nine possible $\left|^{2S+1}L_J\right\rangle$ states that are usually used in potential models of glueballs. Actually, it is rather logical that the helicity formalism reduce to a usual $LS$-basis when particles with spin are considered since all the spin projections are allowed. Let us remark that for $J=0$, the two physical states are given by $\left|^1 S_0\right\rangle$ and $\left|^2D_0\right\rangle$. As the $0^{-+}$ state is a pure $\left|^3P_0\right\rangle$ state, the nondegeneracy of the scalar and pseudoscalar glueballs is \textit{de facto} explained when the valence gluons have a spin degree of freedom. 

\section{Potential model}\label{pmodel}
\subsection{Main Hamiltonian}

The construction of two-gluon helicity states presented in Sec.~\ref{hsglu} is based on purely kinematical arguments. In order to compute a glueball mass spectrum, it is necessary to use a particular Hamiltonian which will contain the dynamics of the system. The simplest way of modeling a two-gluon glueball is to use a two-body spinless Salpeter Hamiltonian with a Cornell potential, that is
\begin{equation}\label{H_cornell}
H_0 = 2\sqrt{\vec{p}^{\, 2}} + a_{g}\, r - 3\, \frac{\alpha_s}{r}.
\end{equation}
The kinetic part is the kinetic energy of two spinless massless particles, i.e. the valence gluons for which the spin is neglected at the dominant order. But, the spin symmetry will be taken into account by the use of helicity states, even if Hamiltonian~(\ref{H_cornell}) is spin-independent. 

The potential term has a Cornell shape, that is a linear-plus-Coulomb form. The linear confining term can be seen as the static energy of a flux tube linking the two gluons. The string tension $a_{g}$ can be related to the string tension of a mesonic flux tube, denoted as $\sigma$, by a scaling law $a_{g}={\cal C}\, \sigma$. A typical value for $\sigma$ is about $0.2$~GeV$^2$, and two values of ${\cal C}$ can be found in the literature: either ${\cal C}=9/4$ (Casimir scaling),
or $3/2$ (square root of Casimir scaling). While the $3/2$ factor is commonly found in bag model-inspired approaches \cite{pcas}, the Casimir scaling seems to be favored by more recent effective approaches and by lattice computations~\cite{cas}. We will here assume the Casimir scaling hypothesis, that is ${\cal C}=9/4$. Beside the long range linear potential, the Coulomb term comes from short range interactions: It is the lowest order approximation of the one gluon exchange diagram between two gluons. $\alpha_s$ is then an effective strong coupling constant smaller than 1 and the factor 3 is the color factor associated with a gluon pair in a color singlet. It was shown in Ref.~\cite{glue2} that, starting from the $0^{++}$ glueball mass and wave function as computed in lattice QCD, the inverse problem can be solved, and the equivalent Hamiltonian is compatible with the form~(\ref{H_cornell}). This validates such an Hamiltonian description, at least in the case where the valence gluons have a spin, since in Ref.~\cite{glue2} it is assumed that the scalar glueball is a $|^1 S_0\rangle$ state.  

Relativistic corrections to the Cornell potential can also be computed. For example, in the flux tube model, corrections to the linear potential appear as a dynamical term proportional to $a_{g}\, \vec L^{\, 2}$ and a spin-orbit term proportional to $a_{g}\, \vec L\cdot\vec S$~\cite{buis07_so}. A nonperturbative retardation term has also been proposed~\cite{ret}. Moreover, relativistic corrections to the Coulomb term can be computed from the QCD Feynman diagrams involving two gluons at tree level. Their complete expression can be found in Ref.~\cite{oge}, and involves the usual contact (spin-spin), spin-orbit, and tensor interactions. For our purpose, it is sufficient just to list the global structure of all these additional terms. All the matrix elements appearing in these tree level relativistic corrections can be found in Tables~\ref{tab1} for helicity states. Finally, it is worth mentioning that the first relativistic corrections are, in this formalism, proportional to $1/\mu^2$, $\mu$ being the dynamical mass $\mu=\left\langle \sqrt{\vec p^{\, 2}}\right\rangle$ gained by the valence gluons because of confinement. 

\subsection{Instanton-induced forces}\label{ins}

It is known that, in the light meson sector, nonperturbative contributions due to instantons comes into play. Roughly speaking, instantons are classical solutions of the Euclidean equations of motion of QCD, which provide informations on the nontrivial vacuum structure of this theory. We refer the reader to Ref.~\cite{instanton3} for a review on instantons in QCD. It has been shown that instantons-induced forces exist between the quark and the antiquark in a light meson. Such forces can be included in potential quark models as an isospin-dependent contact term which is nonzero in the pseudoscalar channel ($0^{-+}$) only~\cite{instanton}. In particular, the strong attractive nature of the instanton-induced contribution in the case of the pion is able to explain the particularly low value of its mass, without lowering the masses of the other mesons. 

If the instanton-induced forces are rather well understood in the meson sector of QCD, the situation is not so clear for glueballs. It has firstly been shown in Ref.~\cite{inst} that instantons induce a strong attractive force in the scalar glueball channel and a repulsive force in the pseudoscalar channel. In the tensor ($2^{++}$) channel, these forces vanish; moreover, instantons are not expected to play any role in the other channels. From the results of a more recent study~\cite{inst2}, it is tempting to assume that the instanton-induced forces in the scalar and pseudoscalar channels are of equal magnitude but of opposite sign. This could be a consequence of the self-duality of the instanton's field strength and a general characteristic of the instanton contribution in all hadrons \cite{Kochelev:2005vd}. From this discussion, and although its exact form has not been computed yet, we can propose the following ansatz for the instanton-induced contribution:
\begin{equation}\label{hidef}
	\Delta H_I=-P\, {\cal I}\, \delta_{J,0}.
\end{equation}
Such a term only contributes for $J=0$ and depends on the parity $P$. Its magnitude is related to the unknown parameter ${\cal I}$, that we assume to be positive and constant in first approximation. 
 
Why is it so interesting to study the influence of instantons on our model? The problem actually comes from the important mass splitting between the $0^{++}$ and $0^{-+}$ glueballs that is observed in lattice QCD. If the valence gluons are spin-1 particles, instanton-induced interactions are not needed to explain this nondegeneracy, as it has been argued in Sec.~\ref{gspin}. But, if the valence gluons are helicity-1 particles, then an additional mechanism is required to lift the mass degeneracy of the scalar and pseudoscalar glueballs. As it can be seen in Table~\ref{tab1}, no correction involving the usually used operators will be able to do that, since their matrix elements are identical for the $0^{++}$ and $0^{-+}$ glueballs. That is why instanton-induced forces are particularly interesting. First, they act in the correct way, increasing the $0^{-+}$ mass and decreasing the $0^{++}$ one. Second, they have already proved to be very useful in the meson case, and, since we know from Refs.~\cite{inst2,instanton3} that instantons play a role in glueballs too, it would seem more coherent if their effects were included in a glueball model also.       

\section{Mass spectrum}\label{numapp}
 
\subsection{Parameters}

Before performing explicit computations, it is necessary to fix the different parameters of our model. As we already said, we focus on valence gluons with vanishing current mass and we assume the Casimir scaling hypothesis: $a_g=(9/4)\sigma$. We set $\sigma=0.185$~GeV$^2$ for the mesonic string tension. This value is located in a rather standard interval: $\sigma\in[0.17,0.2]$~GeV$^2$ is commonly found in the literature. Moreover, this particular value has already given very good results in a previous computation of quarkonium mass spectra in the flux tube model~\cite{buisnew}. Two models will be proposed following that the valence gluons are assumed to be spin-1 (Model A) or helicity-1 (Model B) particles. In Model A, the $0^{++}$ ground state is a $L=S=0$ one. In this case, we have shown in Ref.~\cite{glue2} that $\alpha_s=0.2$ was compatible with the current lattice QCD data. No instanton-induced interaction is needed in this case since the scalar and pseudoscalar glueballs are \textit{de facto} nondegenerate. We thus set ${\cal I}=0$. In Model B, the situation is more similar to the Coulomb gauge approach of Ref.~\cite{szcz}, since gluons with helicity are used. We will take $\alpha_s=0.45$, a value close to the one of Refs.~\cite{cg2,szcz}. In this case, instanton-induced interactions are required, and we set ${\cal I}=0.45$~GeV in order to reproduce at best the $0^{++}$ and $0^{-+}$ masses which are computed in lattice QCD \cite{lat0,lat1}. The parameters appearing in both models are summed up in Table~\ref{tab3}. 

\begin{table}[ht]
\caption{Parameters used in our computations.}
\label{tab3}
\begin{ruledtabular}
\begin{tabular}{lcc}
& Model A & Model B\\
\hline	
$m_g$ & 0 & 0\\     
$\sigma$ (GeV$^2$) & 0.185 & 0.185  \\
$\alpha_s$ & 0.200 & 0.450\\
${\cal I}$ (GeV)& 0 & 0.450\\
\end{tabular}
\end{ruledtabular}
\end{table}

The glueball mass spectrum now remains to be numerically computed from the central spin-independent Hamiltonian $H_0$. Only the radial wave function will be affected by the Hamiltonian of the system, the spin and angular parts being fixed thanks to the helicity formalism. We will use the Lagrange mesh method to compute the matrix elements of $H_0$. This method allows a very simple and accurate treatment of semirelativistic Hamiltonian of the form~(\ref{H_cornell}). We refer the reader to Refs.~\cite{lag1,lag2} for more informations about the Lagrange mesh method. Knowing the matrix representation of $H_0$, the eigenequation 
\begin{equation}\label{eigen}
	 H_0\left|J^{PC}\right\rangle=M_0\left|J^{PC}\right\rangle
\end{equation}
can be solved, and the total mass is given by
\begin{eqnarray}\label{mform}
	M&=&M_0-P\, {\cal I}\, \delta_{J,0}.
\end{eqnarray}

\subsection{Results}

\begin{table}[t]
\caption{Available data for $C=+$ glueball masses from various lattice QCD models and Coulomb gauge QCD (CGQCD), compared with the results of Models A and B. The glueball mass is given and the corresponding spin/helicity state is detailed in both cases. Parameters of Table~\ref{tab3} are used, and all masses are given in GeV.}
\label{tab4}
\begin{ruledtabular}
\begin{tabular}{cllcclcl}
   $J^{PC}$ & Lattice     &Lattice~\cite{lat3}   & CGQCD \cite{szcz} & \multicolumn{2}{c}{Model A} & \multicolumn{2}{c}{Model B}\\
\hline
 $0^{++}$  & 1.710$\pm0.050\pm0.080$~\cite{lat1}& 1.475$\pm0.030\pm0.065$ & 1.980 & 1.655 &$\left|^1S_0\right\rangle$   & 1.724 &$\left|S_+;0^+\right\rangle$\\
           & 2.670$\pm0.180\pm0.130$~\cite{lat0}& 2.755$\pm0.070\pm0.120$ & 3.260 & 2.696 &$\left|^1S_0\right\rangle$   & 2.543 &$\left|S_+;0^+\right\rangle$\\
           &                                    & 3.370$\pm0.100\pm0.150$ &       & 3.101 & $\left|^5D_0\right\rangle$  & 3.234      &$\left|S_+;0^+\right\rangle$ \\
           &                                    & 3.990$\pm0.210\pm0.180$ &       & 3.496     & $\left|^1S_0\right\rangle$ &3.839 &     $\left|S_+;0^+\right\rangle$ \\
 $0^{-+}$  & 2.560$\pm0.035\pm0.120$~\cite{lat1}& 2.250$\pm0.060\pm0.100$ & 2.220 & 2.500 &$\left|^3P_0\right\rangle$ & 2.624 &$\left|S_-;0^-\right\rangle$\\
           & 3.640$\pm0.060\pm0.180$~\cite{lat0}& 3.370$\pm0.150\pm0.150$ & 3.430 & 3.305 &$\left|^3P_0\right\rangle$ & 3.443 &$\left|S_-;0^-\right\rangle$\\
 $1^{-+}$   &                                    &                         &      & 2.500 &$\left|^3P_1\right\rangle$ & Forbidden & \\      
 $1^{++}$   &                                    &                         &      & 3.101 &$\left|^5D_1\right\rangle$ & Forbidden & \\             
 $2^{++}$  & 2.390$\pm0.030\pm0.120$~\cite{lat1} & 2.150$\pm0.030\pm0.100$& 2.420 & 1.655 &$\left|^5S_2\right\rangle$ & 2.588 &$\left|D_+;2^+\right\rangle$\\
           &                                     & 2.880$\pm0.100\pm0.130$& 3.110 & 2.696 &$\left|^5S_2\right\rangle$   & 3.077 &$\left|S_+;2^+\right\rangle$\\
           &                              &      &       & 3.101 &$\left|^{1,5}D_2\right\rangle$   & 3.325&$\left|D_+;2^+\right\rangle$\\ 
 $2^{-+}$  & 3.040$\pm0.040\pm0.150$~\cite{lat1} & 2.780$\pm0.050\pm0.130$& 3.090 & 2.500 &$\left|^3P_2\right\rangle$ & 3.077&$\left|S_-;2^-\right\rangle$ \\
           & 3.890$\pm0.040\pm0.190$~\cite{lat1} & 3.480$\pm0.140\pm0.160$& 4.130 & 3.304 &$\left|^3P_2\right\rangle$ & 3.732&$\left|S_-;2^-\right\rangle$ \\
 $3^{++}$  & 3.670$\pm0.050\pm0.180$~\cite{lat1} & 3.385$\pm0.090\pm0.150$& 3.330 & 3.101 &$\left|^5D_3\right\rangle$ & 3.254 &$\left|D_-;3^+\right\rangle$\\
           &                                   & & 4.290 & 3.783 &$\left|^5D_3\right\rangle$ & 3.882 &$\left|D_-;3^+\right\rangle$\\
$3^{-+}$   &                                     &                        &       & 3.601 & $\left|^3F_3\right\rangle$ & Forbidden & \\           
 $4^{++}$  & 3.650$\pm0.060\pm0.180$~\cite{lat2} & 3.640$\pm0.090\pm0.160$& 3.990 & 3.101 &$\left|^5D_4\right\rangle$ & 3.768 &$\left|D_+;4^+\right\rangle$\\
           &                                  &  & 4.280 & 3.784 &$\left|^5D_4\right\rangle$   & 3.961 &$\left|S_+;4^+\right\rangle$\\
           &                                  &  &       & 4.038 &$\left|^{1,5}G_4\right\rangle$   & 4.328 &$\left|D_+;4^+\right\rangle$\\
 $4^{-+}$  &                                  &  & 4.270 & 3.601 &$\left|^3F_4\right\rangle$ & 3.961 &$\left|S_-;4^-\right\rangle$\\
           &                                  &  & 4.980 & 4.204 &$\left|^3F_4\right\rangle$ & 4.499&$\left|S_-;4^-\right\rangle$ \\
 $5^{++}$  &                                  &  &       & 4.038 &$\left|^5G_5\right\rangle$ & 4.207 &$\left|D_-;5^+\right\rangle$\\
 $5^{-+}$  &                                  &  &       & 4.432 &$\left|^3H_5\right\rangle$ & Forbidden &\\
 $6^{++}$  &  & 4.360$\pm0.260\pm0.200$  &       & 4.038&$\left|^5G_6\right\rangle$ & 4.598&$\left|D_+;6^+\right\rangle$\\ 
           &                           &         &       & 4.585 &$\left|^5G_6\right\rangle$   & 4.708&$\left|S_+;6^+\right\rangle$ \\
           &                            &        &       & 4.793 &$\left|^{1,5}I_6\right\rangle$   & 5.073&$\left|D_+;6^+\right\rangle$ \\
\end{tabular}
\end{ruledtabular}
\end{table}

\begin{figure}[ht]
\includegraphics*[width=10cm]{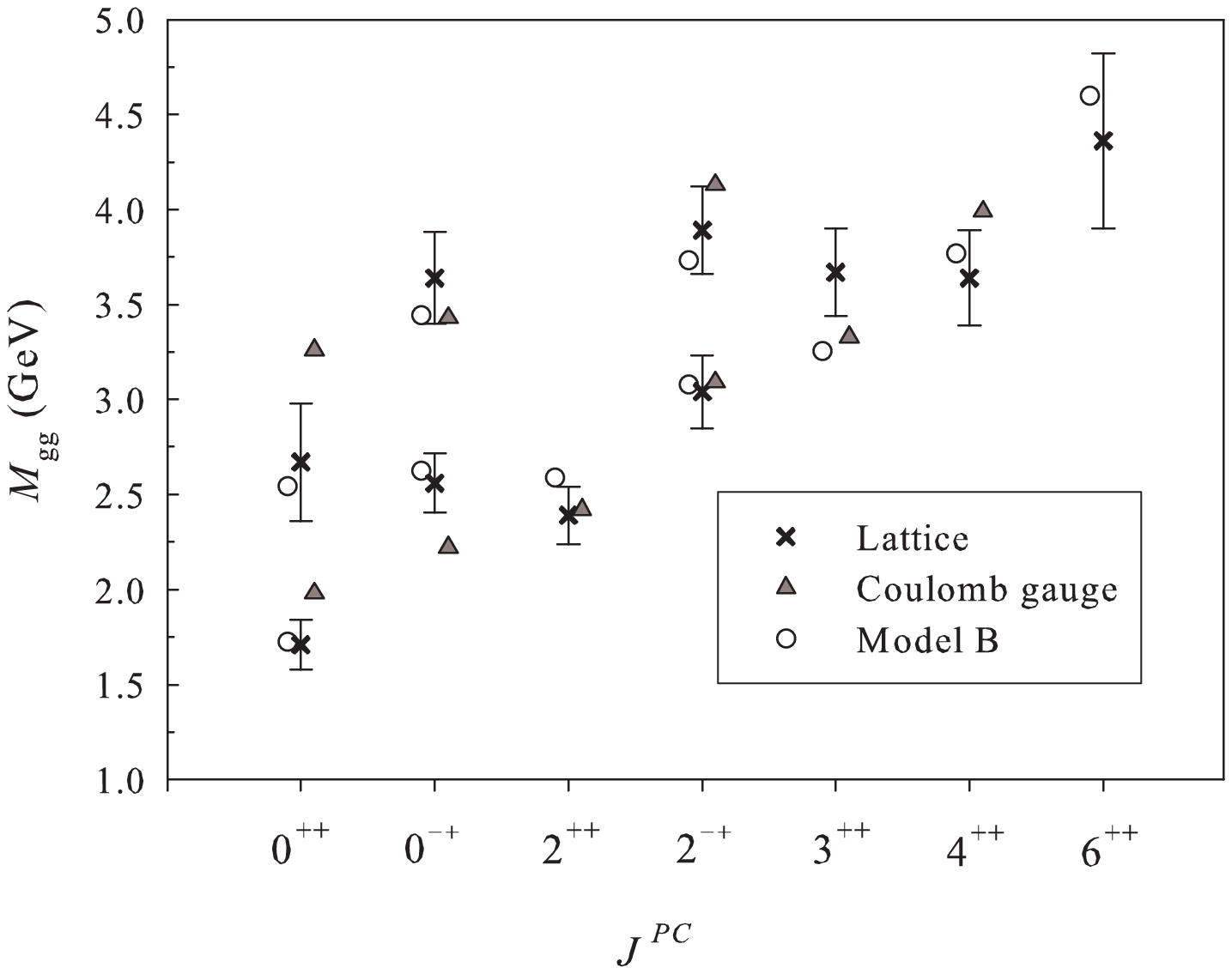}
\caption{Comparison between the lattice QCD data concerning $C=+$ glueballs (crosses), the Coulomb gauge results (triangles)~\cite{szcz}, and our Model B (circles). Masses are given in GeV. All lattice data come from Refs.~\cite{lat0,lat1}, except data for $4^{++}$ and $6^{++}$ states \cite{lat3} (see 2nd and 3rd columns of Table~\ref{tab4}).}
\label{Fig1}
\end{figure}

A glueball mass spectrum can now be computed within the framework of our spinless Salpeter model with the helicity formalism. The results will be compared to some lattice QCD predictions concerning the glueball masses (see Table~\ref{tab4} and Fig.~\ref{Fig1}). The parameters of our models are fitted on data taken from Ref.~\cite{lat1} completed by data from Refs.~\cite{lat0,lat2}. Results from Ref.~\cite{lat3} are also given. The predictions of this reference are quite different from the results coming from the compilation of Refs.~\cite{lat0,lat1}. The ground states of Ref.~\cite{lat3} have lower masses, and more excited states have been computed.  

It is also interesting to compare our results with the Coulomb gauge approach of Ref.~\cite{szcz}. In this last reference, helicity-1 gluons are considered and encoded in the Coulomb gauge formalism. The main features of the mass spectrum of Ref.~\cite{szcz} should then be similar to our Model B. This is roughly the case but with a serious exception: The mass gap between $0^{++}$ and $0^{-+}$ states is about $200$~MeV, which is far less than the value predicted by lattice calculations. But no instanton-induced interaction is explicitly taken into account in this work.

A detailed glueball spectrum is given in Table~\ref{tab4} for Models A and B. We computed the masses of more states than those which are currently observed in lattice QCD. Some of them have a mass greater than 4~GeV; glueball spectrum in lattice QCD is poorly known above this energy range. However, there are higher $0^{++}$ and $2^{++}$ states that lie under this limit with both sets of parameters. Some of them are seen in Ref.~\cite{lat3} but not in Refs.~\cite{lat0,lat1}. It should be interesting to know whether future lattice computations will confirm or not the existence of these states. We also point out again that no $J=1$ state is present at low energy as expected from lattice QCD with helicity-1 gluons. 

Let us begin by a discussion of the results obtained with Model A. In this case, the scalar and pseudoscalar glueballs are compatible with lattice QCD without invoking instanton-induced interactions. But, the situation gets clearly worse for higher $J$. First, $J=1$ states are present, which are not observed in lattice QCD. In particular, the rather light $1^{-+}$ glueball seems to be a serious flaw of Model A since the $-+$ channel is rather well known from lattice QCD in this energy range, and no such state has been seen. Then, the lightest $2^{++}$ state is degenerate with the $0^{++}$ state, and it should not be the case. Actually, nearly every state with Model A does not lie within the error bars of lattice QCD, suggesting that this model should be discarded.    
   
We turn now our attention to Model B. In this case, an instanton-induced term is needed, otherwise the $0^{++}$ and $0^{-+}$ glueball would have the same mass. The value ${\cal I}=0.450$~GeV is of the same order than the typical magnitude of instanton-induced effects in mesons~\cite{instanton3}. Globally, the rest of the spectrum is in agreement with lattice QCD, and the agreement is far better than with Model A. Our $(2k+2)^{++}$ states, although being roughly compatible with lattice QCD, lie in the upper part of the errors bars, while the $3^{++}$ state is too light. A modification of the parameters $\sigma$ or $\alpha_s$ would shift the whole spectrum in the same direction. So an improvement of the $3^{++}$ mass would spoil the rest of the spectrum. Spin-dependent interactions at tree level (see Table~\ref{tab1}) are the same for $\left|D_+;J^+\right\rangle$ and  $\left|D_-;J^+\right\rangle$. So, these interactions would shift the $3^{++}$ and $2^{++}$ masses in the same direction. We suggest that additional mechanism such as singlet-doublet mixing -- corrections beyond tree level -- could cure this mass problem. 
Let us note that, in Ref.~\cite{szcz}, even though the $2^{++}$ is located inside the lattice error bars, the $3^{++}$ is also below. Model B is thus rather convincing, particularly because the number of states is drastically decreased with helicity-1 gluons: The few states that are observed in lattice QCD are the only one that can be built, without extra states as it is the case in Model A. The mass spectrum of Model B has been plotted in Fig.~\ref{Fig1} and is compared to lattice QCD and Coulomb gauge data.  

\begin{table}[t]
\caption{
Comparison of $C=+$ glueball mass ratios, normalized to the lightest
$0^{++}$ state, between the lattice QCD data~\cite{lat0,lat1}, and our Model B (the corresponding helicity state is given). Parameters of Table~\ref{tab3} are used. The lattice mass ratios followed by~\cite{lat0} are taken from Table VIII of this last reference. The other ones, that were not given in Ref.~\cite{lat0}, have been computed with the more recent data of Ref.~\cite{lat1}.}
\label{tab5}
\centering
\setlength{\extrarowheight}{5pt}
\begin{ruledtabular}
\begin{tabular}{clcl}
   $J^{PC}$ & Lattice   & \multicolumn{2}{c}{Model B}\\
\hline
$0^{++}$ & 1                         & 1.00 &$\left|S_+;0^+\right\rangle$\\
         & 1.54$\pm$0.11~\cite{lat0} & 1.48 &$\left|S_+;0^+\right\rangle$\\
$0^{-+}$ & 1.50$\pm$0.04~\cite{lat0} & 1.52 &$\left|S_-;0^-\right\rangle$\\
         & 2.11$\pm$0.06~\cite{lat0} & 2.00 &$\left|S_-;0^-\right\rangle$\\
$2^{++}$ & 1.39$\pm$0.04~\cite{lat0} & 1.50 &$\left|D_+;2^+\right\rangle$\\
$2^{-+}$ & 1.79$\pm$0.05~\cite{lat0} & 1.78 &$\left|S_-;2^-\right\rangle$ \\
         & 2.27$\pm$0.09~\cite{lat1} & 2.16 &$\left|S_-;2^-\right\rangle$ \\
$3^{++}$ & 2.15$\pm$0.09~\cite{lat1} & 1.89 &$\left|D_-;3^+\right\rangle$\\
\end{tabular}
\end{ruledtabular}
\end{table}

\begin{figure}[ht]
\includegraphics*[width=10cm]{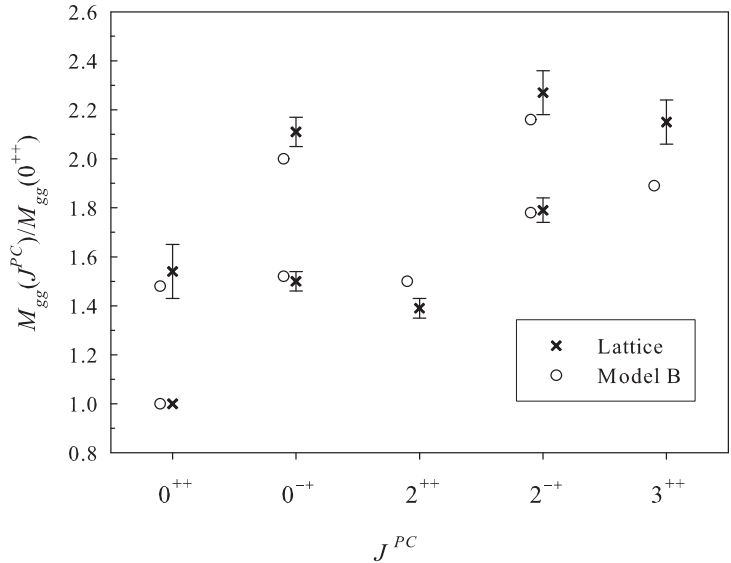}
\caption{Comparison of $C=+$ glueball mass ratios, normalized to the lightest
$0^{++}$ state, between the lattice QCD data (crosses)~\cite{lat0,lat1}, and our Models B (circles). More details on these data can be found in Tables~\ref{tab4} and \ref{tab5}.}
\label{Fig2}
\end{figure}

The errors of lattice data on absolute glueball masses are quite large. This is due the the uncertainty in setting the mass scale. This problem can be corrected by computing mass ratios. We choose to report all masses to the lowest $0^{++}$ state as in Ref.~\cite[Table VIII]{lat0}. This state does not present the lowest statistical uncertainty but it is the ground state. The mass ratios for our model B, computed with data from Table~\ref{tab4}, are given in Table~\ref{tab5} and Fig.~\ref{Fig2}, and are compared with the corresponding lattice data~\cite{lat0,lat1}. The study of mass ratios brings the same conclusions as the discussion above about absolute masses. The mass ratios obtained with Model A are not mentioned since this model does not bring relevant results. 

\section{Conclusions}\label{conclu}

We have presented in this work an application of the helicity formalism developed in Ref.~\cite{jaco} for bound states of two valence gluons. We have shown that this helicity formalism only leads to the quantum numbers which are observed in lattice QCD. This feature is a considerable improvement of usual potential models in which lots of extra glueball states are obtained. In particular, the well-known absence of $J=1$ states has been shown to be a consequence of the symmetrization in the helicity formalism for helicity-1 gluons. $J=1$ states are however allowed for spin-1 gluons. 

We have then developed a simple potential model, relying on a spinless Salpeter Hamiltonian with a Cornell potential and massless gluons. However, we stress that our framework, based on the helicity formalism, can be applied to any existing potential model provided that the helicity states are used to compute the different matrix elements. An extra physical mechanisms has been considered: instanton-induced interactions, for which we have proposed a phenomenological term taking into account their main properties. Using the Lagrange mesh method, the glueball spectrum coming from our Hamiltonian model can be computed within the helicity formalism. Two possibilities were taken into account: Either the valence gluons are spin-1 or helicity-1 particles. We have computed the masses of different states by using standard values for the string tension and the strong coupling constant. The instanton parameter has been fitted so that it leads to an optimal agreement with lattice QCD data in the scalar and pseudoscalar channels. It appears that, if gluons are spin-1 particles, no instanton-induced term is needed in the scalar and pseudoscalar channels, but the rest of the spectrum is not in agreement with lattice QCD. If helicity-1 gluons are assumed, an instanton term has to be added since the $0^{\pm+}$ glueballs would be nondegenerate otherwise. Its value is similar to the one encountered in the meson sector. In this case, the glueball mass spectrum is in good agreement with lattice QCD, both qualitatively (no extra $J^{P+}$ states are obtained) and quantitatively. 

By comparing the results of both pictures with lattice QCD, it appears that modeling the currently known $C=+$ glueballs by a bound state of two massless valence gluons with helicity-1 seems to be far more relevant, justifying \textit{a posteriori} the assumption that two-gluon states dominate in this sector. In particular, the necessity of adding instanton-induced forces should not be seen as a flaw of the model, but rather as a way to be more coherent with other studies showing that instantons contribute in glueballs. 

In conclusion, the helicity formalism appears to be a very promising way to improve potential models of hadrons containing valence gluons, because it takes into account correctly the relativistic character of these particles. It is relevant to assume that such valence gluons are massless particles with helicity-1, as we argued in this paper. For what concerns the glueball mass spectrum, it is remarkable how the addition of the helicity formalism to a simple potential model as the one we developed here leads to such a nice agreement with lattice QCD. In future works, we plan to present a more accurate glueball model, still within the helicity formalism but including relativistic corrections. 

\acknowledgments
F. B. and C. S. thank the F.R.S.-FNRS for financial support. V. M. thanks the IISN for financial support. The authors are grateful to N. Boulanger and T. Schaefer for valuable discussions and advice about the present work.

\end{document}